\documentclass{aa}  

\usepackage{graphicx}
\usepackage{txfonts}
\usepackage[english]{babel}
\usepackage{amssymb}
\usepackage{amsmath}
\usepackage{mathrsfs}
\usepackage{latexsym}
\usepackage{longtable}
\usepackage{multirow}
\usepackage{epsf}
\usepackage{color}
\usepackage{float}
\usepackage{enumerate}
\usepackage{txfonts}
\usepackage{natbib}
\usepackage{comment}
\usepackage{capt-of}
\usepackage{rotating}
\usepackage{ulem}

\newcommand{\hd}{HD\,100453 }
\newcommand{ \s}{$\sim$}
\newcommand{\dg}{$^\circ$}

\begin{document}

   \title{Shadows and spirals in the protoplanetary disk \hd \thanks{Based on observations performed with SPHERE/VLT under program ID 096.C-0248(B)}}

   \author{M.~Benisty\inst{1}
   \and
   T.~Stolker\inst{2}
    \and
   A.~Pohl\inst{3}
     \and
   J.~de~Boer\inst{4}
      \and
G.~Lesur\inst{1}
\and 
C.~Dominik\inst{2}
\and
C.\,P.~Dullemond\inst{5}
\and
M.~Langlois\inst{6,19}
\and
M.~Min\inst{7,2}
\and
K.~Wagner\inst{8}
\and
T.~Henning\inst{3}
\and
A.~Juhasz\inst{9}
\and
P.~Pinilla\inst{4,8}
\and 
S.~Facchini\inst{10}
\and
D.~Apai\inst{8}
\and
R.~van Boekel\inst{3}
\and 
A.~Garufi\inst{11,12}
\and 
C.~Ginski\inst{4}
\and
F.~M\'enard\inst{1}
\and
C.~Pinte\inst{1}
\and
S.~P.~Quanz\inst{12}
\and
A.~Zurlo\inst{13,14,19}
\and
A.~Boccaletti\inst{15}
\and M.~Bonnefoy\inst{1}
\and  J.L.~Beuzit\inst{1}
\and G.~Chauvin\inst{1}
\and M.~Cudel\inst{1}
\and S.~Desidera\inst{16}
\and M.~Feldt\inst{3}
\and C.~Fontanive\inst{17}
\and R.~Gratton\inst{16}
\and M.~Kasper\inst{18,1}
\and A.-M.~Lagrange\inst{1}
\and H.~LeCoroller\inst{19}
\and D.~Mouillet\inst{1}
\and D.~Mesa\inst{16}
\and E.~Sissa\inst{16}
\and A.~Vigan\inst{19} 
\and J.~Antichi\inst{20}
\and T.~Buey\inst{13}
\and T.~Fusco\inst{21,19}
\and D.~Gisler\inst{12}
\and M.~Llored\inst{18}
\and Y.~Magnard\inst{1}
\and O.~Moeller-Nilsson\inst{3}
\and J.~Pragt\inst{2}
\and R.~Roelfsema\inst{2}
\and J.-F.~Sauvage\inst{21,19}
\and F.~Wildi\inst{22}
}

  \institute{Univ. Grenoble Alpes, CNRS, IPAG, F-38000 Grenoble, France.  \email{Myriam.Benisty@univ-grenoble-alpes.fr}
               \and Anton Pannekoek Institute for Astronomy, University of Amsterdam, Science Park 904, 1098 XH Amsterdam, The Netherlands
                \and Max Planck Institute for Astronomy, K\"{o}nigstuhl 17, 69117 Heidelberg, Germany
		\and Leiden Observatory, Leiden University, P.O. Box 9513, 2300 RA Leiden, The Netherlands
                  \and Institute for Theoretical Astrophysics, Heidelberg University, Albert-Ueberle-Strasse 2, D-69120 Heidelberg, Germany
                  \and CRAL, UMR 5574, CNRS, Universit\'{e} Lyon 1, 9 avenue Charles Andr\'{e}, 69561 Saint Genis Laval Cedex, France
                  \and SRON Netherlands Institute for Space Research, Sorbonnelaan 2, 3584 CA, Utrecht, The Netherlands
                  \and Department of Astronomy/Steward Observatory, The University of Arizona, 933 North Cherry Avenue, Tucson, AZ 85721, USA
                  \and Institute of Astronomy, Madingley Road, Cambridge CB3 OHA, UK
                  \and Max-Planck-Institut fur Extraterrestrische Physik, Giessenbachstrasse 1, 85748 Garching, Germany
                  \and Universidad Auton\'onoma de Madrid, Dpto. F\'isica Te\'orica, Facultad de Ciencias, Campus de Cantoblanco, E-28049 Madrid, Spain
                  \and Institute for Astronomy, ETH Zurich, Wolfgang-Pauli-Strasse 27, 8093 Zurich, Switzerland
                  \and N\'ucleo de Astronom\'ia, Facultad de Ingenier\'ia, Universidad Diego Portales, Av. Ejercito 441, Santiago, Chile 
                  \and Departamento de Astronom\'ia, Universidad de Chile, Casilla 36-D, Santiago, Chile 
                  \and LESIA, Observatoire de Paris-Meudon, CNRS, Universit\'{e} Pierre et Marie Curie, Universit\'{e} Paris Didierot, 5 Place Jules Janssen, F-92195 Meudon, France
                \and INAF-Osservatorio Astronomico di Padova, Vicolo dell'Osservatorio 5, Padova, Italy, 35122-I 
                \and Institute for Astronomy, University of Edinburgh, Blackford Hill View, Edinburgh EH9 3HJ, UK 	      
                \and European Southern Observatory, Karl-Schwarzschild-Str. 2, D85748 Garching, Germany 
                \and Aix Marseille Univ, CNRS, LAM, Laboratoire d'Astrophysique de Marseille, Marseille, France 
\and INAF - Osservatorio Astrofisico di Arcetri, Largo E. Fermi 5, 50125 Firenze, Italy
	\and ONERA - 29 avenue de la Division Leclerc, F-92322 Chatillon Cedex, France
	\and Geneva Observatory, Univ. of Geneva, Chemin des Maillettes 51, 1290 Versoix, Switzerland
}

                \date{}

  \abstract
   {Understanding the diversity of planets requires studying the morphology and  physical conditions in the protoplanetary disks in which they form.} 
   {We aim to study the structure of the  \s10\,Myr old protoplanetary disk HD~100453, to detect features that can trace disk evolution and to understand the mechanisms that drive these features.}
   {We observed  \hd in polarized scattered light with SPHERE/VLT at optical (0.6\,$\mu$m, 0.8\,$\mu$m) and near-infrared (1.2\,$\mu$m) wavelengths, reaching an  angular resolution of \s0.02\arcsec{}, and an inner working angle of \s 0.09\arcsec{}. }
   {We spatially resolve the disk around HD~100453, and detect polarized scattered light up to \s0.42\arcsec{} (\s48\,au). We detect a cavity, a rim with  azimuthal brightness variations at an inclination of \s38\dg~with respect to our line of sight, two shadows and two symmetric spiral arms. The spiral arms originate near the location of the shadows, close to the semi major axis. We detect a faint feature in the SW that can be interpreted as the scattering surface of the bottom side of the disk, if the disk is tidally truncated by the M-dwarf companion currently seen at a projected distance of \s119\,au.  We construct a radiative transfer model that accounts for the main characteristics of the features with an inner and outer disk misaligned by $\sim$72$^\circ$. The azimuthal brightness variations along the rim are well reproduced with the scattering phase function of the model.  While  spirals can be triggered by the tidal interaction with the companion, the close proximity of the spirals to the shadows suggests that the shadows could also play a role. The change in stellar illumination along the rim  induces an azimuthal variation of the scale height that can contribute to the brightness variations. }
   {Dark regions in polarized images of transition disks are now detected in a handful of disks and often interpreted as shadows due to a misaligned inner disk. However, the origin of such a misalignment in HD\,100453, and of the spirals, is still unclear, and might be due to a yet-undetected massive companion inside the cavity, and on an inclined orbit. Observations over a few years will allow us to measure the spiral pattern speed, and determine if the shadows are fixed or moving, which may constrain their origin. }


   \maketitle

\section{Introduction}
Thousands of exoplanetary systems have been detected so far displaying a wide diversity in their architecture.  Understanding planet formation and its outcomes requires good knowledge of the protoplanetary disks at different spatial scales. Although forming planets have not been unambiguously detected  so far, one can aim to study the conditions for their formation by looking for indirect signatures and imprints of the mechanisms driving the disk evolution.

In  recent years, high resolution images of protoplanetary disks have shown a variety of small-scale features. In the sub-millimeter regime, one of the most stunning images was obtained using ALMA at its highest angular resolution, on HL Tau, a very young object \citep[0.5\,Myrs old,][]{HLTau}, and revealed concentric rings in a very flat disk \citep{carrasco2016, pinte2016}. These rings indicated that planet formation might occur very early in the disk lifetime, but alternative explanations, such as hydrodynamical instabilities were also proposed \citep{flock2015, ruge2016, bethune2016}. Interestingly, rings were also detected in the sub-millimeter observations of a very old disk, TW\,Hya \citep[10\,Myrs old;][]{andrews2016,tsukagoshi2016} suggesting that such small features are ubiquitous and/or long lived. On the other hand, other sub-millimeter images showed, azimuthally asymmetric brightness enhancements in  continuum \citep{casassus2012, marel2015, perez2014, pinilla2015a} and in very few objects, spiral arms \citep[][P\'erez et al., in press]{christiaens2014}. 
The diversity of these features supports the idea that several processes (e.g., planet formation, hydrodynamical instabilities, photoevaporation) might act simultaneously and with different relative contribution depending on the object.    

Stunning images of the scattering surfaces of protoplanetary disks are produced with polarimetric differential imaging \citep[PDI; e.g.][]{kuhn2001,apai2004,quanz2011}. The technique consists of measuring the linear polarization of the light scattered by dust grains in the disk and to remove the unpolarized contribution, including that from the star. Recent images show rings \citep[e.g.][]{rapson2015,wolff2016,ginski2016}, spiral arms \citep[e.g.,][]{muto2012, grady2013, benisty2015, stolker2016}, localized dips  \citep[e.g.,][]{pinilla2015b, canovas2016} and shadows \citep[e.g.,][]{avenhaus2014}.  As these observations only trace  small dust grains in the upper disk layers, and not the bulk of the disk mass, these features may trace enhancements in surface density, or variations in the disk scale height due to  local heating events \citep{juhasz2015, pohl2015}.  
These features have now been observed in disks surrounding stars with a broad range of properties in terms of stellar luminosity,  age and disk evolution. 

Of particular interest for this paper is \hd A, hereafter referred to as simply HD\,100453, a Herbig A9Ve star  located in the Lower Centaurus Association \citep{kouwenhoven2005}, at \s114$^{+11}_{-4}$\,pc (Perryman et al. 1997), with an early-M star companion \citep{chen2006}. In a detailed multi-wavelength study, \citet{collins2009} refined the age of the system to be 10$\pm$2\,Myr, and also constrained the companion properties. It is an M4.0 - M4.5V, 0.20$\pm$0.04\,M$_\odot$ star, located at 1.045\arcsec{}$\pm$0.025\arcsec{} (i.e., \s119\,au) at a PA of 126$\pm$1$^\circ$.   \hd was classified as a Group~I (flared) disk by \citet{meeus2001}. The disk reprocesses a significant fraction of the stellar light in the inner and outer disk regions suggesting a vertically thick and flared disk \citep{dominik2003}. Interestingly,  there is no clear sign of accretion onto the star.  \citet{collins2009} derived an accretion rate upper limit of 1.4$\times$10$^{-9}$ M$_\odot$/yr from the FUV continuum, confirmed by \citet{fairlamb2015} (upper limit of 4.9 10$^{-9}$ M$_\odot$/yr). \hd gas tracers also show a peculiarity: while Herbig stars with a strong NIR excess show 4.7\,$\mu$m CO emission \citep{brittain2007}, \hd does not show any \citep{collins2009}, which suggests a high dust-to-gas ratio  or a reduction of the gas content in the inner disk.  \citet{collins2009} report a non-detection of CO J=3-2 with the JCMT, that indicates that the gas amount in the outer disk region might also be severely reduced.  From the 1.2\,mm continuum emission, and the CO upper limit, the disk mass is estimated to be 8$\times$10$^{-5}$ M$_\odot$, and the gas to dust ratio to be not more than 4:1 \citep{collins2009}. 

The disk surrounding \hd must be relatively compact, compared to other Herbig Ae disks. HST observations report no scattered light detection beyond 3" \citep{collins2009}. A background star is detected at a projected distance of 90\,au, which indicates that the disk is either truncated by tidal interaction with the M-dwarf companion, or optically thin, at this projected distance from the star. This is supported by two marginally resolved images, at \s0.2\arcsec{}-0.3\arcsec{} scales (i.e., \s25-35\,au), in the PAH and Q-band filters \citep{habart2006,khalafinejad2016}.  Using SPHERE with differential imaging, \citet{wagner2015} reported the detection of two spiral arms in scattered light, up to 0.37\arcsec{} (\s42\,au), and a marginal detection of a gap or cavity inside 0.18\arcsec{} (\s20\,au).  

In this paper, we report the first polarized differential images of \hd obtained in the optical ($R'$ and $I'$ bands) and in the near infrared ($J$ band) with SPHERE/VLT. The paper is organized as follows. Section 2 describes the observations and the data processing. Section 3 reports on the detected disk features, Sect.~4 provides a radiative transfer model that well reproduces the observations and in Sect.~5 we discuss our findings.




\begin{figure*}
	\centering
	\begin{tabular}{c}
        \includegraphics[width=0.7\textwidth]{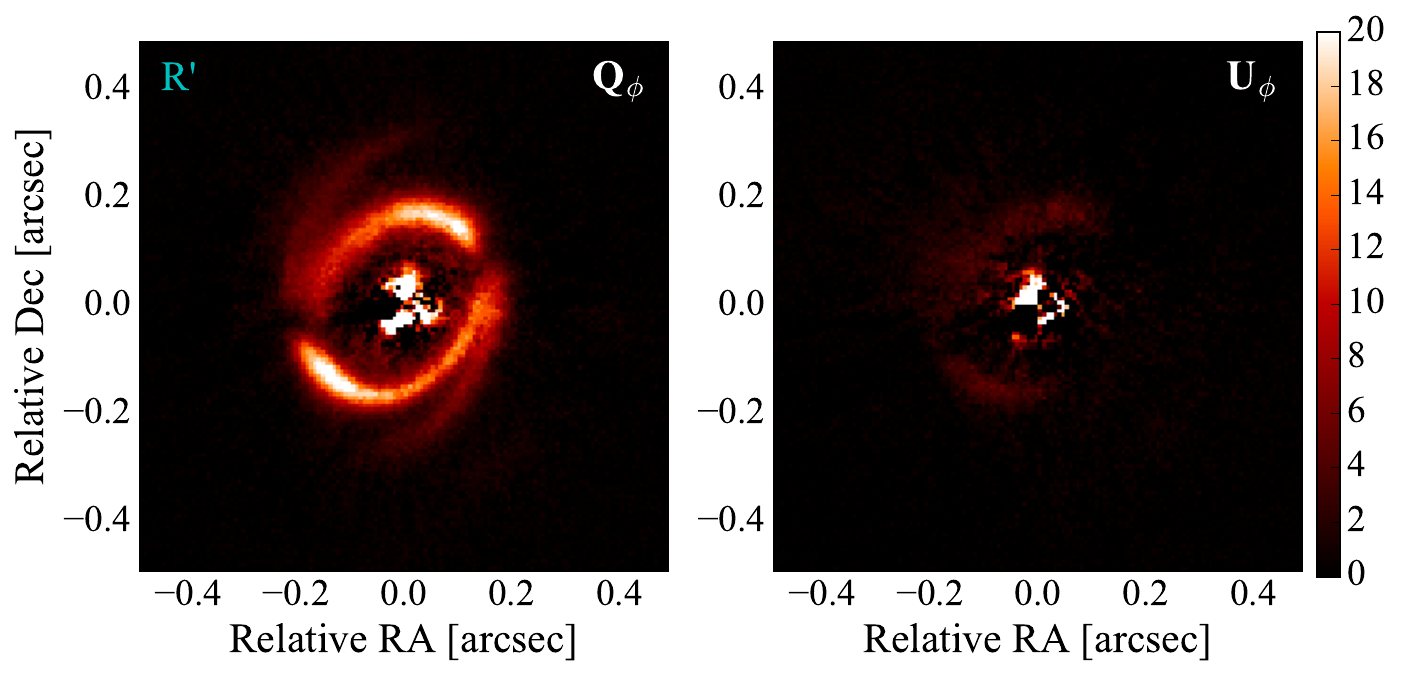}\\	
          \includegraphics[width=0.7\textwidth]{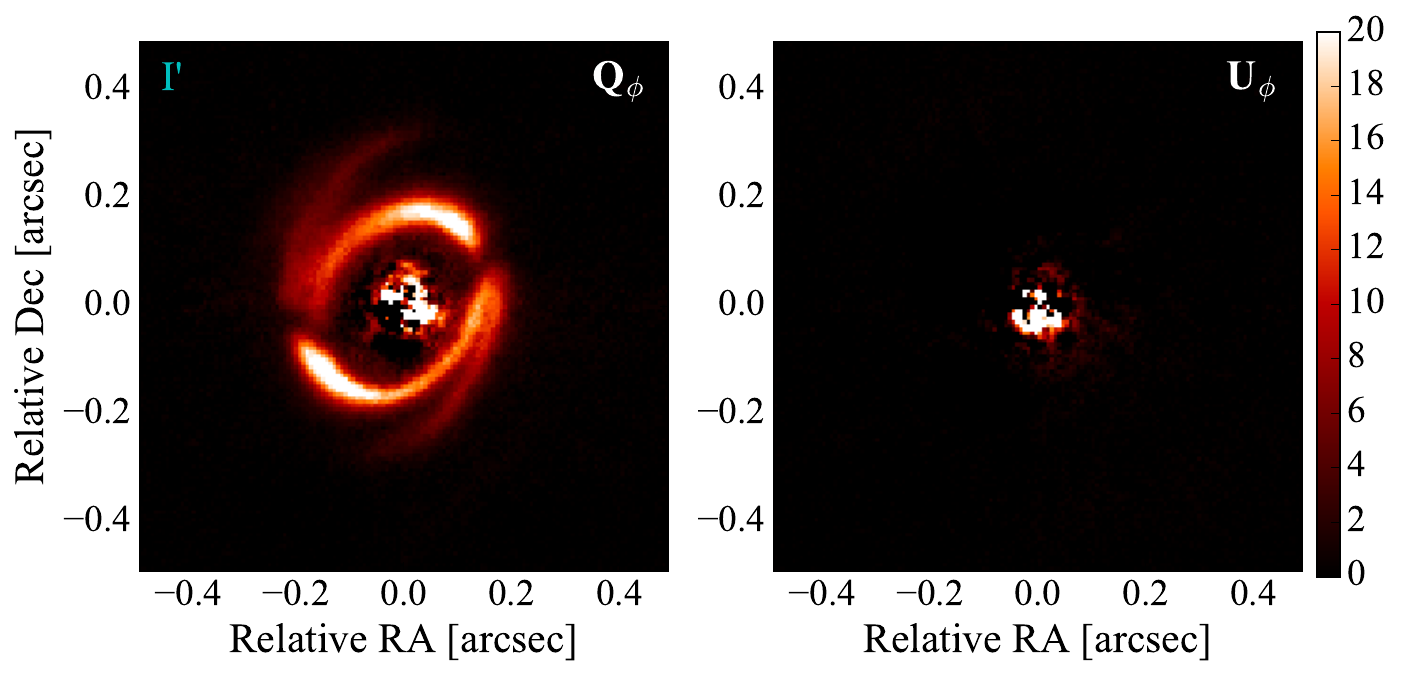}\\	
	\includegraphics[width=0.7\textwidth]{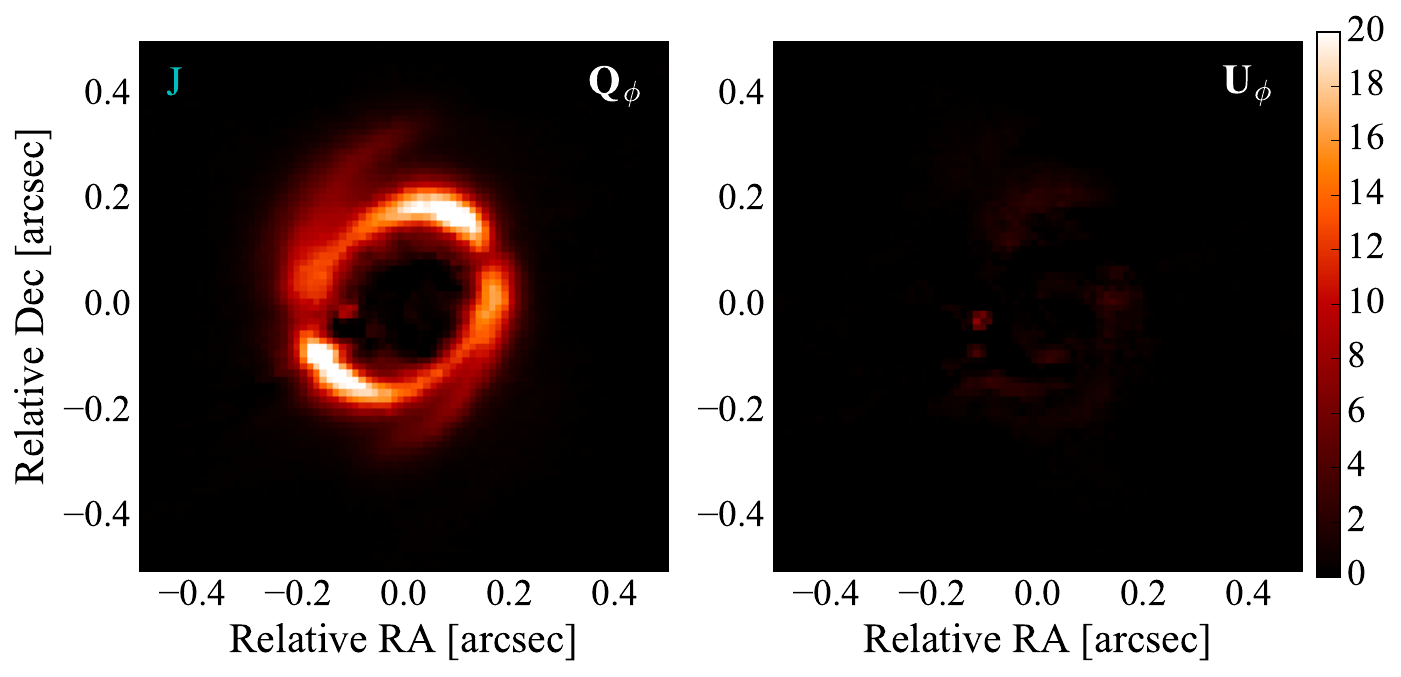} 
	\end{tabular}
	\caption{$R'$ (top), $I'$ (middle), and $J$ band (bottom) polarized intensity images, $Q_{\phi}$ (left) and $U_{\phi}$ (right). In the optical images, the inner bright region corresponds to saturated pixels inside our IWA. In the NIR images, the inner dark region is masked by the coronagraph. The color scale of the $Q_{\phi}$ and $U_{\phi}$ are the same, and arbitrary. For all images, East is pointing left.}
	\label{fig:images}
\end{figure*}

\section{Observations and data reduction}
\label{sec:obs}
The observations were carried out on March 30th and 31th, 2016, with the SPHERE instrument \citep[][Beuzit et al. in prep.]{sphere2008}, equipped with an extreme adaptive-optics (AO) system \citep{fusco2006,petit2014,sauvage2014}  at the Very Large Telescope at Cerro Paranal, Chile. The observations were executed through the Guaranteed Time program. \hd was observed in the $R'$ and $I'$ band filters ($\lambda_0$=0.626, $\Delta\lambda$=0.149~$\mu$m; $\lambda_0$=0.790, $\Delta\lambda$=0.153~$\mu$m, respectively) using the ZIMPOL instrument \citep{zimpol2010,thalmann2008} with a plate scale of 3.5~milli-arcseconds (mas) per pixel and in the $J$ band ($\lambda_0$=1.258, $\Delta\lambda$=0.197~$\mu$m) using the infrared dual-band imager and spectrograph  \citep[IRDIS;][]{dohlen2008, maud2014}, with a plate scale of 12.25~mas per pixel.

We used a 185~mas-diameter coronographic focal mask combined with an apodized pupil and Lyot stop. \hd was observed for 85 and 80 minutes with IRDIS and ZIMPOL, respectively. These data were taken under moderate  AO  conditions (seeing between 0.7 and 1.0\arcsec{}). 
From an analysis of the reference point spread function (PSF), we find that the AO quality reaches a diffraction-limited regime, with a 20.8$\times$24 mas resolution (slightly elongated PSF due to wind speed, the theoretical diffraction limit being 20.6\,mas) and a 43\% Strehl Ratio at 0.8\,$\mu$m.

For polarimetric differential imaging, the instruments split the beam into two orthogonal polarization states.  The half-wave plate (HWP) that controls the orientation of the polarization, and allows to decrease the effect of instrumental polarization, was set to four positions shifted by 22.5$^\circ$ in order to construct a set of linear Stokes vectors.  The data was reduced according to the double difference method \citep{kuhn2001}, which is described in detail for the polarimetric modes of IRDIS and ZIMPOL in \citet{deboer2016}, and lead to the Stokes parameters $Q$ and $U$. Under the assumption of single scattering, the scattered light from a circumstellar disk is expected to be linearly polarized in the azimuthal direction. Hence, we describe the polarization vector field in polar rather than Cartesian coordinates \citep{avenhaus2014} and define the
polar-coordinate Stokes parameters $Q_\Phi$ and $U_\Phi$ as:

\begin{equation}
Q_\Phi = +Q \cos(2\Phi) + U \sin(2\Phi), 
\end{equation}
and
\begin{equation}
U_\Phi = -Q \sin(2\Phi) + U \cos(2\Phi), 
\end{equation}

\noindent where $\Phi$ is the position angle of the location of interest (x, y) with respect to the  star location (x$_{0}$,y$_{0}$), and is written as: 
\begin{equation}
\Phi = \arctan\frac{x - x_{0}}{y-y_{0}} + \theta.
\end{equation}

$\theta$ corrects for instrumental effects such as the angular misalignment of the HWP. In this coordinate system, the azimuthally polarized flux from a circumstellar disk appears as a consistently positive signal in the $Q_\phi$ image, whereas the $U_\phi$ image remains free of disk
signal and provides a convenient estimate of the residual noise in the $Q_\phi$ image  \citep{schmid2006}.  
To determine the absolute disk surface brightness in polarized intensity requires advanced calibration of the polarimetric throughput of the system, which lies beyond the scope of this study. We therefore use arbitrary units in the images shown in the paper.

In Fig.~\ref{fig:images}, we present the resulting polarized scattered light images in the optical and NIR.  We note that there is a residual signal in the  $U_\phi$ image, in particular in the $R'$ band image, that may be due to multiple scattering events \citep{canovas2015}. 


\section{Polarized intensity images}
\label{sec:images}

The images of Fig.~\ref{fig:images} reveal a number of disk features. The NIR image shows the same features as the optical ones, albeit with a lower angular resolution, leading to fuzzier features. 

Looking at the optical images (Fig.\ref{fig:images}, top and middle), beyond a distance of 0.09\arcsec{} (\s10\,au) that corresponds to the inner working angle (IWA) of our observations, we detect, from inside out: 

\textbf{(a)} a region with low scattered light signal, called \textit{cavity}, from our IWA up to \s0.14\arcsec{} (\s16\,au).  We note that although we can not probe inside 0.09\arcsec{}, the NIR excess seen in the SED indicates the presence of a significant amount of dust grains in the inner au(s). The inner working angle therefore provides an upper limit on the outer radius of the inner disk.

\textbf{(b)}~a ring-like feature, called the \textit{rim}, located at \s0.14\arcsec{} (\s16\,au) with an apparent width ranging from \s0.050 to \s0.075\arcsec{} (\s 5 to 9\,au). Its brightness varies azimuthally, and there are two clear maxima at PAs  \s135\dg~and \s325\dg. The brightest regions are distributed over an azimuthal range of \s70\dg. 

\textbf{(c)}~two dark regions along the rim, that we refer to as \textit{shadows}. These regions are located at \s100\dg~and \s293\dg~and have an angular extent of \s12\dg~at the inner edge of the rim, that slightly increases with radius.

\textbf{(d)}~two spiral arms, in the NE and the SW, extending to \s0.42\arcsec{} (\s48\,au) and \s0.34\arcsec{} (\s39\,au), respectively. Interestingly, the spirals are located very close to the shadows.

\begin{figure}
	\centering
	\includegraphics[width=0.5\textwidth]{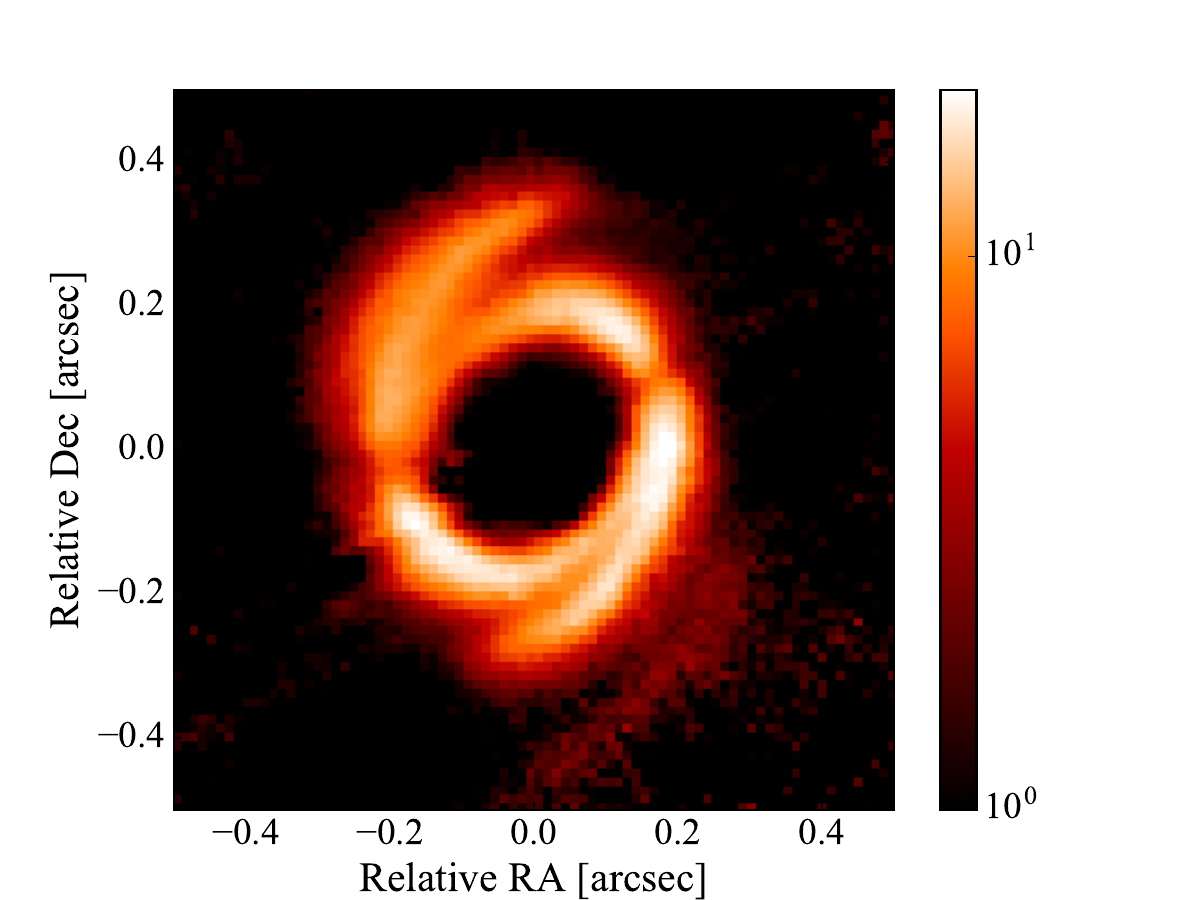}
	\caption{$J$-band $Q_{\phi}$ image after scaling each pixel by the square of its distance from the star, $r^{2}$. The scaling takes into account the geometry of the $\tau$=1 surface given by our radiative transfer model. The   color  log-scale is arbitrary. }
	\label{fig:third}
\end{figure}

\textbf{(e)}~an additional spiral-like feature, in the SW. This feature can be seen in Fig.~\ref{fig:third}, in which we scale the  $J$-band image by r$^2$ to compensate for the $r^{-2}$ dependency of the stellar illumination, and enhance faint features located further out in the disk.  

The values of r$^2$  applied to the original image take into account the inclination and PA of the object, as well as the disk flaring following the method described by \citet{stolker2016b} (and using a $\tau=1$ surface with $h=0.22\times r^{1.04}$ with r and h in au, as derived from our radiative transfer model, see Sect.~\ref{sec:RT}). This feature is also detected in the ZIMPOL data,  in the differential imaging data by \citet{wagner2015} and in newly acquired angular differential images with SPHERE (see Fig.~\ref{fig:wagner}).


In all images, the NE spiral appears to have a larger opening angle than the SW spiral. If we assume that the disk is inclined and flared, and that the spirals intrinsic opening angles are similar, this may indicate that the NE is the far side of the disk and the SW its near side.  This is supported by the smaller width of the rim in the SW and of the shadow in the West, and by the fact that the SW spiral is twice as bright as the NE spiral in the total intensity images of \citet{wagner2015}, assuming that this effect is due to forward-scattering. Finally, if the disk is truncated by the M-dwarf, the faint additional spiral-like feature in the SW may be tracing scattered light from the outer edge of the bottom side of the disk. Assuming that the images in Fig.\,\ref{fig:images} show signal from the disk surface layer at a given height \s$h$ from the disk midplane, this additional spiral-like feature would trace the layer at \s$-h$, on the other side of the disk midplane. This  scenario would support the idea that the SW is indeed the near side of the disk.

\begin{figure}
	\centering
	\includegraphics[width=0.5\textwidth]{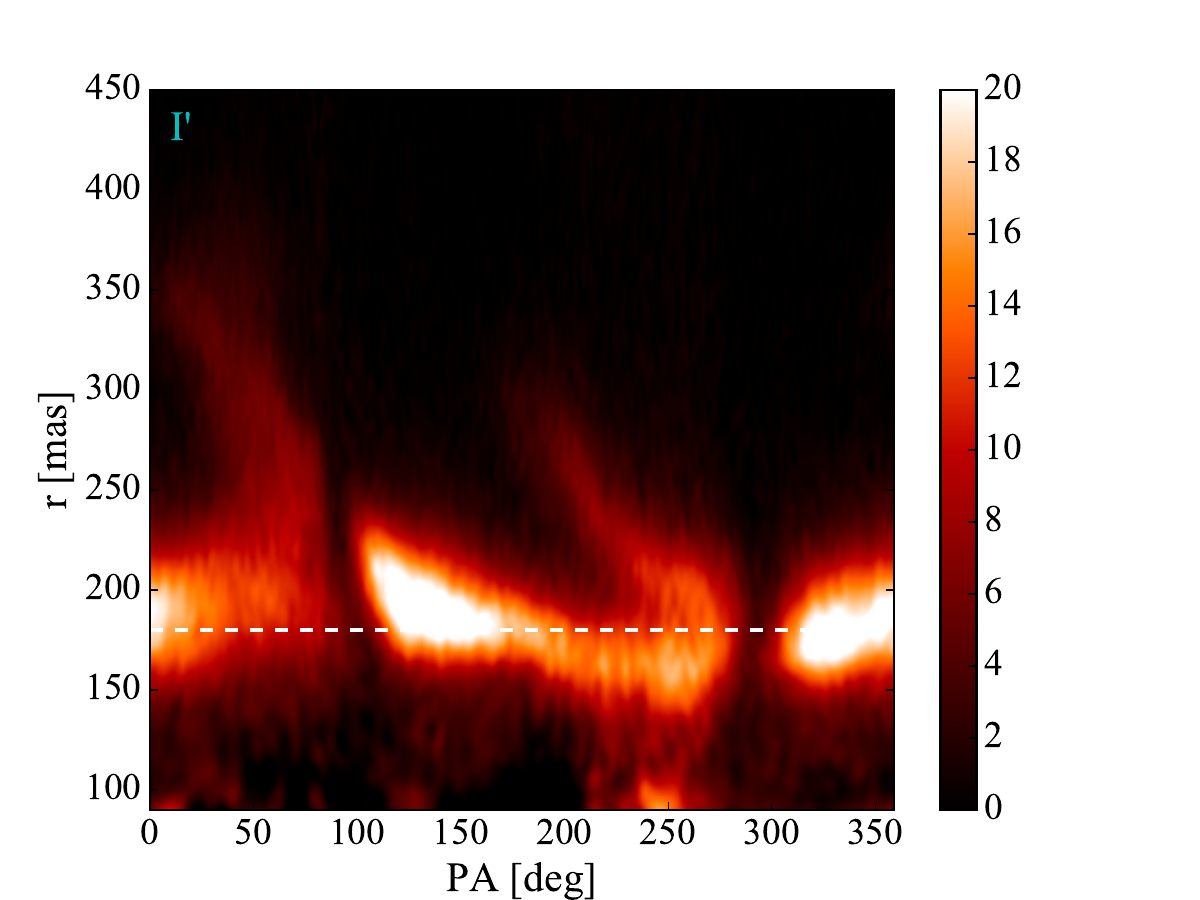}
	\caption{Polar map of the deprojected $I'$  $Q_{\phi}$ image using $i=38^\circ$ and $PA=142^\circ$. The dashed line indicates a radius of 0.18\arcsec{}.  The color scale is arbitrary. }
	\label{fig:polar}
\end{figure}

To determine the inclination and the position angle of the rim, we fit an ellipse to the brightest point along each radius in the optical image, and find major and minor axes corresponding to an inclination of \s38\dg, in close agreement with the value found by \citet{wagner2015} (34\dg). A position angle of \s142\dg\, and a shift of the ellipse center by \s7\,mas in the SE fit the data best. This offset could originate from either the vertical thickness of the inclined rim (assumed to be zero in our 2D-ellipse fitting), a non-zero eccentricity of the rim, an effect of the dust grain scattering phase function, or a combination of some or all of these.


In Fig.~\ref{fig:polar} we present the polar mapping of the $I'$  image, after deprojection using i=38\dg\,and PA=142\dg. It clearly shows the shadows and the azimuthal brightness variations. Interestingly, the NE spiral seems to appear on both sides of the shadow (at approximatively PA\s100\dg).

\begin{figure*}
	\centering
	\begin{tabular}{cc}
	\includegraphics[width=0.45\textwidth]{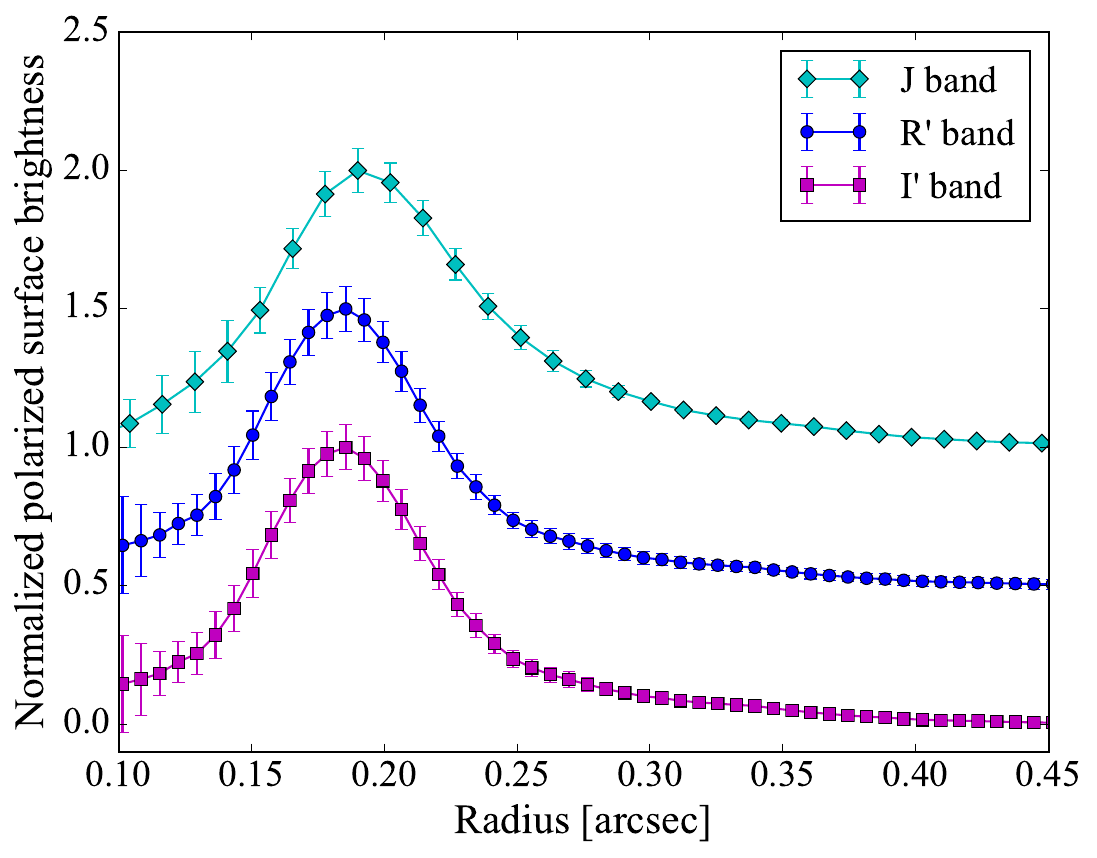} & 
	\includegraphics[width=0.44\textwidth]{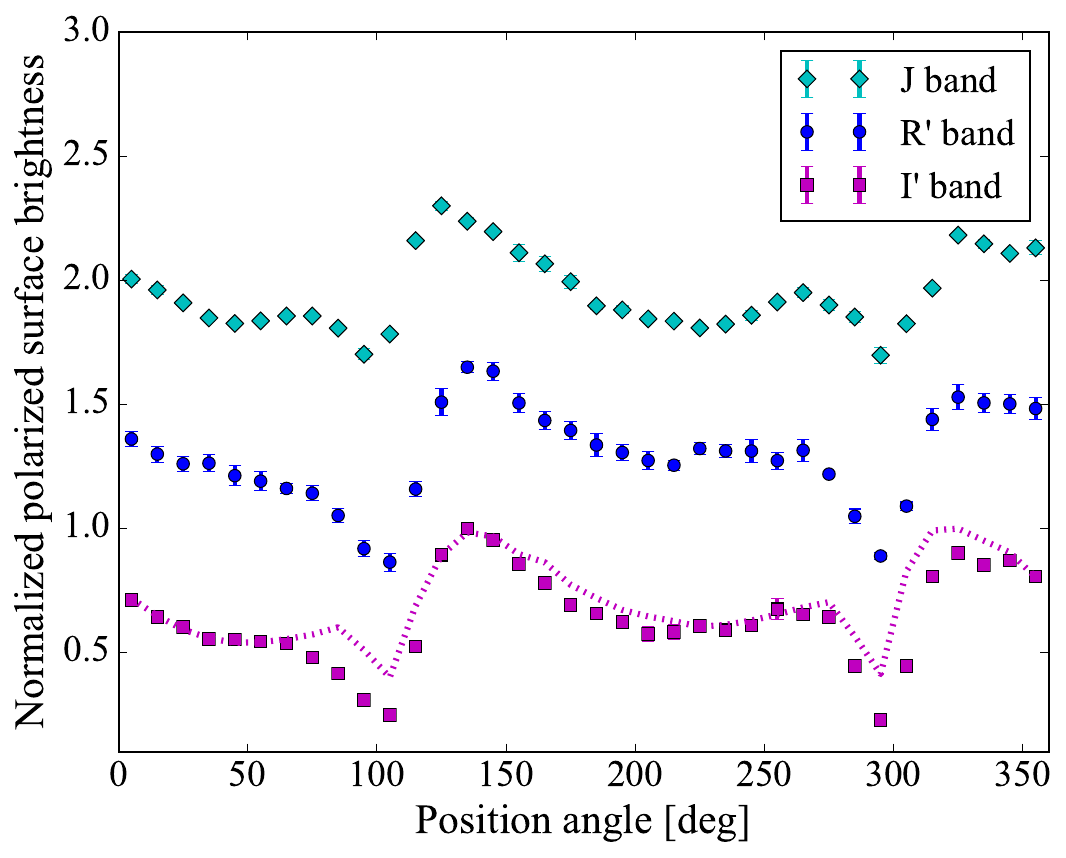}  
	\end{tabular}
	\caption{Left: Normalized radial cuts of the $R'$, $I'$ and $J$-band images after averaging azimuthally. Right: Normalized azimuthal cuts of the $R'$, $I'$ and $J$-band images, after averaging radially between 170 and 200 mas. The radiative transfer model prediction (dotted curve) reproduces the observed azimuthal brightness variations relatively well. Note that due to the large variation of surface brightness along the rim, the standard deviation in each bin can vary from 2 to 17\%. The curves are shifted vertically for clarity.} 
	\label{fig:cut}
\end{figure*}

Figure~\ref{fig:cut} shows the radial and azimuthal cuts, when averaging azimuthally and across the rim width (0.17\arcsec{}-0.20\arcsec{}), respectively. The error bars are estimated as the standard deviation in each bin in the  $U_\phi$ image. We measure a ratio of (radially averaged) polarized surface brightness of \s5 between the shadows and the brightest regions of the rim.  





\section{Radiative transfer modeling}
\label{sec:RT}
In this section, we aim to provide a radiative transfer model for HD\,100453, that reproduces the main characteristics of the rim, the spirals and the shadows seen in the scattered light images, in particular, their locations, widths, and brightness variations. 


\subsection{MCMax3D model}
We use the 3D version of the continuum radiative transfer code MCMax \citep{min2009} which calculates the thermal structure of the disk and produces ray-traced images. We consider an inner disk, a cavity and an outer disk. The dust surface density is parametrized radially as:
\begin{equation}
\label{eq:surface_daensity}
\Sigma(r) \propto r^{-\epsilon} \exp{\left[ -\left(\frac{r}{R_{\rm tap}}\right)^{2-\epsilon} \right]},
\end{equation}
where $R_{\rm in} < r < R_{\rm out}$ is the disk radius, $R_{\rm tap}$ the tapering-off radius and $\epsilon$ the surface density power law index. The surface density profile is scaled to the total dust mass, M$_{\rm dust}$ and the vertical density distribution follows a Gaussian profile.  The disk aspect ratio is parametrized radially as $H(r)/r = (H_0/r_0) (r/r_0)^\psi$ with $H(r)$ being the scale height, $H_0/r_0$  the aspect ratio at the reference radius $r_0$, and $\psi$  the flaring index.  

We consider a minimum ($a_{\rm min}$) and maximum  ($a_{\rm max}$) grain size and use a power law for the dust grain size distribution with an index $\gamma$. We use a dust mixture made of 70\% silicates and 30\% carbon \citep[DIANA,][]{woitke2016}, and the porosity of the grains is set to 25\%. 
We consider the grains to be irregular in shape by setting the maximum volume void fraction used for the distribution of hollow spheres (DHS) method to 0.8 \citep{min2005}.

We describe the spiral arms as Archimedean spirals, following $r(\theta) = A1 + A2*(\theta-\theta_0)^n$. We assume that they trace perturbations in the disk scale height, rather than in the surface density. Hence, along the spirals, the scale height is multiplied by 1+ $a_{\rm{height}} * exp^{((r-r(\theta))/w)^2}*(A1/r)^q$, where w is the width of the spiral and q determines the steepness of the radial falloff of the spiral arm. 

Once the temperature structure is computed, synthetic SEDs and ray-traced polarized images can be produced at any wavelength. We compute monochromatic Stokes $Q_{\phi}$ and $U_{\phi}$ images, at 0.79\,$\mu$m, and use an unsaturated observed Stokes $I$ frame as a PSF to convolve the synthetic maps.

\begin{figure}
	\centering
	\includegraphics[width=0.55\textwidth]{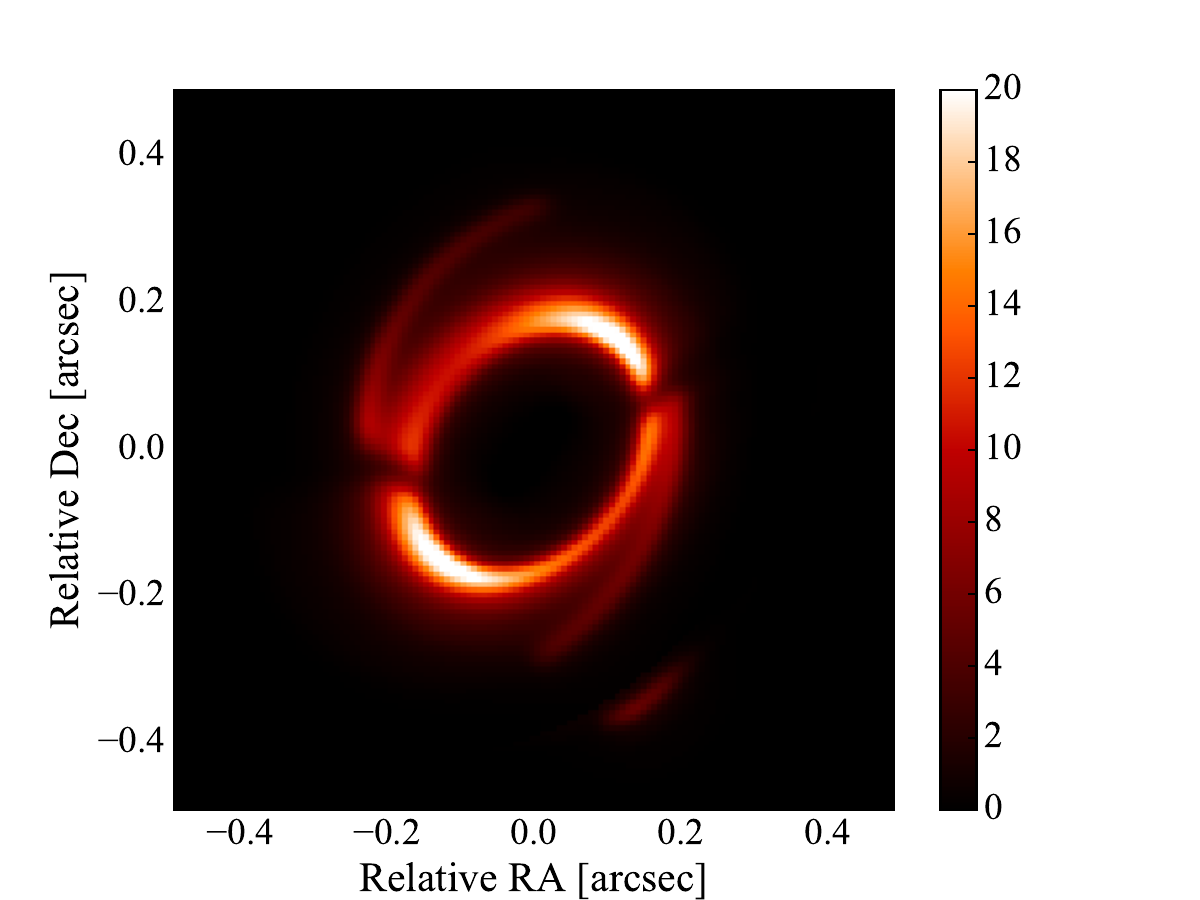} 
	\caption{Synthetic $Q_{\phi}$ $I'$-band polarized image from our radiative transfer model. To show it more clearly, the  faint feature in the SW is enhanced by a factor of five, and traces the scattering surface of the bottom side of the disk. }
	\label{fig:rt}
\end{figure}

\begin{figure}
	\centering
		\includegraphics[width=0.45\textwidth]{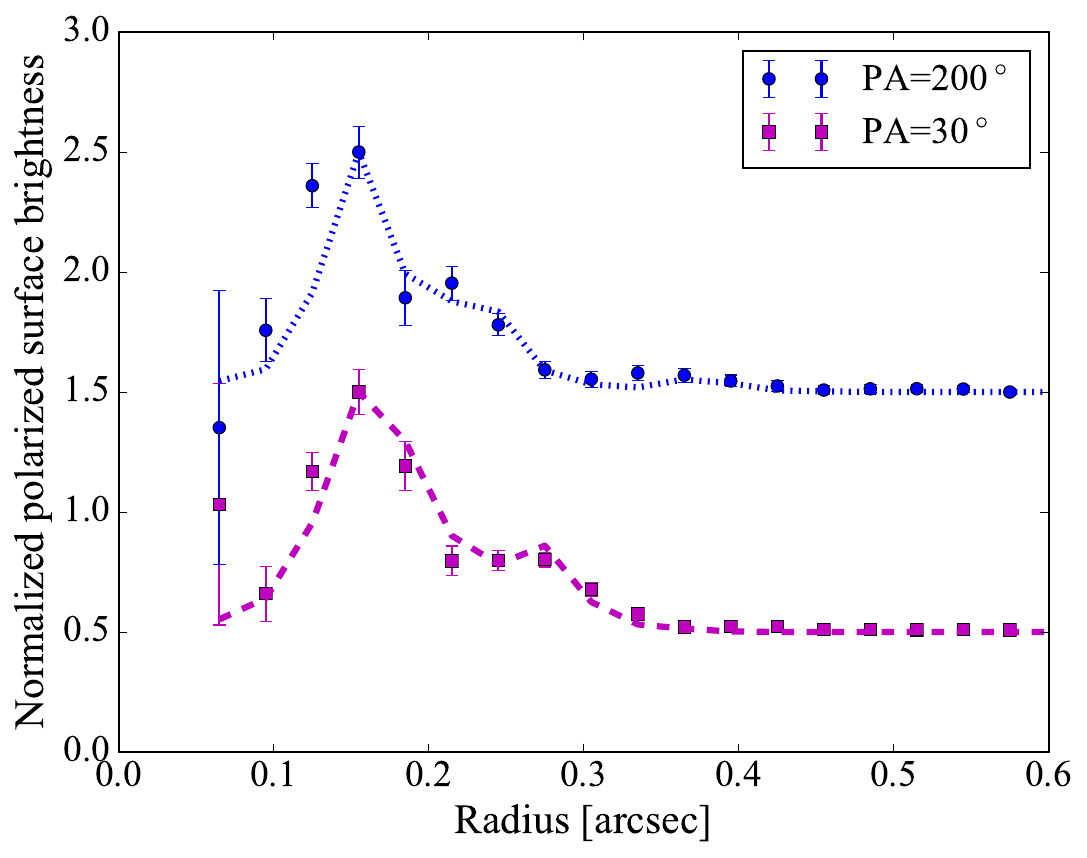}	
	\caption{Normalized radial cuts along position angles of 30\dg\,and 200\dg\,(obtained within a 10\dg\,bin), and the corresponding model predictions (dashed and dotted lines, respectively).  The curves are shifted vertically for clarity.} 
	\label{fig:cutPA}
\end{figure}

\subsection{Best model}
We generate the shadows using a misaligned inner disk, with respect to the outer disk. For the outer disk, we use the inclination and position angles derived from the ellipse fitting (i\s38\dg, PA\s142\dg; see Sect.~\ref{sec:images}), while for the inner disk, we use i\s48\dg, PA\s80\dg, which are obtained from geometrical model fitting of NIR interferometric observations (PIONIER survey, Lazareff et al. subm). As the inner and outer disks must be significantly misaligned to create deep shadows \citep{marino2015}, we assume that the near side of the outer disk is in the SW, while the near side of the inner disk is in the NE. This leads to a misalignment of \s72$^\circ$, obtained by calculating the angle between the normal vectors to the inner and outer disks.  The location of the shadows depends on the orientation of the inner disk (for a given outer disk orientation), while their shape  depends on the inner disk aspect ratio (the larger the aspect ratio, the broader the shadows), and on the width and roundness of the outer disk rim. 

Our model parameters are summarized in Table~\ref{table:model_parameters}.  We fix the inner disk rim at 0.27\,au (Klarmann et al. subm) and its outer radius at 1\,au \citep{menu2015}. The outer disk starts at 20\,au with a strongly peaked surface density profile (see Fig.~\ref{fig:sigmased}, left).  We use a minimum grain size of 0.01\,$\mu$m and a maximum size of 1\,$\mu$m, as larger grains result in a strong brightness asymmetry between the near and far side of the disk, due to forward scattering, which we do  not observe. The outer disk mass, aspect ratio and flaring were chosen to fit the mid IR and far IR excesses in the SED. We note that the grain size distribution that we consider is valid for the surface layers  probed in the scattered light images, but probably not valid for the disk midplane that likely hosts larger grains. 

Figure~\ref{fig:rt} shows our best model that well reproduces  the location of the rim and its azimuthal brightness variation as well as the width and location of the shadows (Fig.~\ref{fig:cut}).  The brightness contrast between the rim and the spirals is also well reproduced (Fig.~\ref{fig:cutPA}). We find opening angles of \s15\dg\,at the onset of the spirals. The surface brightness is maximal on both sides of the major axis due to the high degree of polarization for 90\dg\,scattering angles. A third spiral-like feature in the SW is also predicted by our model, and overlaps with the one detected in the observations, supporting the idea that it could trace the  scattering surface of the  bottom side of the disk, if it is truncated at \s50\,au (Fig.~\ref{fig:rt}). The SED is well reproduced for wavelengths longer than 10 microns, most of which is probing the outer disk, but the model misses significant emission in the NIR  (Fig.~\ref{fig:sigmased}, right). It is a general problem that disk models fail to reproduce the NIR excess of Herbig~Ae stars \citep[e.g.,][Klarmann et al. subm.]{benisty2010, flock2016}, and solving this is beyond the scope of this paper. In the particular case of HD~100453,  \citet{khalafinejad2016} added an optically thin halo to reprocess a significant fraction of the stellar light. Optically thin material at high altitude was similarly considered in models of other Herbig\,Ae stars to reproduce the large NIR excess \citep[e.g.,][]{verhoeff2011,wagner2015a}. 






\begin{table}[t]
\caption{MCMax3D model parameters. For the star, we used the following parameters: T$_{\rm{eff}}$=7400\,K, L=8\,$L_{\odot}$, R=1.73\,$R_{\odot}$, M=1.66\,$M_{\odot}$. Note that we use a negative value for the outer disk inclination to account for the fact that the near side is in the SW.}
\label{table:model_parameters}
\centering
\bgroup
\def\arraystretch{1.25}
\begin{tabular}{c|c|c}
\hline\hline
Parameter & Inner disk & Outer disk\\ 
\hline
R$_{\rm in}$ [au] & 0.27 & 20   \\
R$_{\rm out}$ [au] & 1 & 45 \\ 
R$_{\rm tap}$ [au] & 50 & 50 \\ 
\hline
M$_{\rm dust}$ [M$_\odot$] & $1 \times 10^{-10}$ & $2\times 10^{-5}$ \\
$\epsilon$ & 1 & -3 (rim)\\
 &  & 1 (r$\geq$25\,au)\\

H$_0$/r$_0$ & 0.04 & 0.05 \\
r$_0$ [au] & 1 & 20 \\
$\psi$ & 0 & 0.13 \\
$\alpha$ & $10^{-3}$ & $10^{-3}$ \\
\hline
a$_{\rm min}$ [$\mu$m] & 0.01 & 0.01 \\
a$_{\rm max}$ [$\mu$m] & 1 & 1 \\
$\gamma$ & -3.5 & -3.5 \\
\hline
i [deg] & 48 & -38 \\
PA [deg] & 80 & 142 \\
\\
\hline
\hline
Parameter &  NE spiral & SW spiral \\
\hline
R$_{\rm in}$ [au] & 27 & 38   \\
R$_{\rm out}$ [au] & 33 & 45 \\
A1 [au] & 27 & 30 \\
A2 [au] & 7 & 8 \\
$\theta_0$ [deg] & 125 & 125 \\
n & 1.12 & 1.12 \\
a$_{\rm{height}}$ & 0.8 & 1.1 \\
w [au] & 1.2 & 1.2 \\
q & 1.7 & 1.7 \\
\end{tabular}
\egroup
\end{table}

\section{Discussion}

\subsection{Origin of the spirals}
Until now, spiral arms have been unambiguously detected in PDI observations of six Herbig~Ae disks : \hd \citep[][this work]{wagner2015}; AB\,Aur \citep{hashimoto2011}; HD\,142527 \citep{canovas2013,avenhaus2014}; SAO\,206462 \citep{muto2012, garufi2013, stolker2016}; MWC\,758 \citep{grady2013, benisty2015}, and HD\,100546 \citep{ardila2007,garufi2016}. In half of them, the spiral arms  show an m=2 symmetry.  Since these spirals appear in polarized scattered light, they only trace the small dust grains, well coupled to the gas, but located at the surface layers of the disks. It is difficult to know whether they originate in perturbations in the surface layers only, or if they also trace perturbations deeper in the disk.  
In the sub-millimeter wavelength range, that traces the bulk material of the disk, so far only two of these disks show clear spiral arms in the CO lines \citep{christiaens2014, tang2012}, and only one other in the continuum \citep{perez2016}. 

Various mechanisms have been suggested for the origin of the spirals observed in disks. Planet disk interactions launch spiral waves at the Lindblad resonances \citep[e.g.][]{ogilvie2002}, with small pitch angles, while gravitational instabilities lead to large-scale spiral arms with larger pitch angles \citep[e.g.,][]{lodato2004, pohl2015}, capable of trapping dust particles \citep{dipierro2015}.  Non-ideal magnetohydrodynamics \citep[eg.,][]{lyra2015} and shadows can also induce spirals \citep{montesinos2016}.  While all these processes can possibly act together, gravitational instabilities are unlikely to occur in HD\,100453, considering the low gas content of the disk \citep{collins2009} 
The striking symmetry of the two spiral arms seen in HD\,100453 could be induced by two (yet-undetected) planets located inside the cavity. However, in this scenario, the planets should  be located at symmetrical locations inside the cavity, in an unstable configuration. We find this scenario unlikely, also because the m=2 mode is seen in other objects. 

The two symmetric spiral arms seen in \hd can be induced by the tidal interaction with the low-mass companion located at a projected distance of \s119\,au \citep{dong2016}. We note, however, that to be similar to the observations, the disk model presented in \citet{dong2016} is required to be close to face-on, which is not supported by our observations. As there is possibly a wide range of  disk and orbital parameters that would likely lead to a good agreement with the observations, we cannot rule out this possibility as the origin of the spirals, and still find this scenario  likely. 

If not coincidental, the proximity of both of the spirals to the shadows in the polarized intensity images of \hd suggests that the shadows could also play a role, and that the spirals might be induced by the pressure decrease at the shadows' locations \citep{montesinos2016}. We note however, that the stellar and disk parameters considered in the hydrodynamical simulations of \citet{montesinos2016} are very far from the ones measured for HD\,100453. In particular, the Toomre parameter values for \hd are much higher than the minimum ones (ranging from 0.5 to 3.4) in their simulations, and while it is not clear whether it is relevant for HD\,100453,  self-gravity might play an important role in triggering and maintaining the spirals. This is suggested by the non-stationarity of the spirals (see their Fig.~2), in contrast with the expectations in the case of a steady shadow.  Dedicated hydrodynamical simulations are needed to determine the conditions in which shadow-induced spirals could appear in HD\,100453. 

If spirals can be induced by steady shadows, the cooling timescale is required to be much shorter than the dynamical timescale (\s instantaneous), otherwise the gas does not have time to adjust and the pressure gradient is not significant enough to trigger spirals.  On the other hand, if the inner disk (that we assume is responsible for the shadows) precesses, the shadows are not fixed anymore. At the radius that co-rotates with the shadows, the shadowed gas is maintained in a cold region and the disk undergoes the strongest heating/cooling which might lead to spiral density waves, even with non-instantaneous cooling.  For this to apply to HD\,100453, the precession timescale must equal the orbital timescale at the radius where the spirals originate. We note that at the rim location (\s25\,au), the orbital period is \s100\,years, already relatively fast compared to precession timescales \citep{papaloizou1995}. 

Interestingly, the spirals generated by fixed or moving shadows are different. As in the case of a perturbing planet that co-rotates with the disk, if the shadows move at the precession rate of the inner disk, the spirals are trailing, and the rotational direction of the disk is counterclockwise. In contrast, if the spirals are induced by fixed shadows, the outer spirals are leading, and the rotational direction of the disk is clockwise. Such a difference in the gas kinematics will likely be tested by forthcoming ALMA observations of HD\,100453.

\subsection{Shadows-induced scale height variations} 
\label{sec:1dmodel}
At a given radius while orbiting the star, the gas periodically goes  from an illuminated region, with large irradiation, to one with negligible irradiation heating (the shadow). Assuming that the cooling and heating timescales are shorter than the dynamical (orbital) timescale, the gas temperature and the pressure are lower in these shadowed regions.  As the pressure support of the gas fails, the gas falls towards the midplane, reducing the scale height.  Upon exiting the shadow, the gas is heated again, causing the column to expand vertically again. This modulation of the disk scale height might affect the appearance of the rim in scattered light.  To quantify this effect, we consider a single radius of the rim that is directly illuminated by the star. At this radius, the temperature contrast is the strongest between the rim and the shadowed regions, and we assume that radii in the far reaches of the shadow that receive grazing radiation can be neglected.  We applied Newton's second law of motion to $H$, the pressure scale height. We consider the vertical hydrostatic balance equation in the disk as a starting point and  follow the evolution of a vertical gas parcel along the rim as:
\begin{equation}
	\frac{d^2 H(t)}{dt^2} =  -\underbrace{ \Omega_{\mathrm{K}}^2 H(t)}_{\substack{1}}  + \underbrace{\frac{c_{\mathrm{s}}(t)^2}{H(t)}}_{\substack{2}}  - \underbrace{\Gamma\, \frac{dH(t)}{dt}}_{\substack{3}}\,,
	\label{eq:1d_scaleheight}
\end{equation}


\noindent where $c_{\mathrm{s}}$ is the sound speed, $\Omega_{\mathrm{K}}$ the orbital Keplerian frequency, and $\Gamma$ a damping factor.  This second order equation is similar to that of a driven damped oscillator.  On the right hand side of Eq.~\ref{eq:1d_scaleheight},  (1) describes  the vertical component of the gravitational force that tries to contract the disk,  (2) is the vertical pressure force that intends to expand the disk and (3) is a damping term, that mimics the loss of energy. $\Gamma$ is used to characterize the strength of the damping force and is assumed to be on the order of the dynamical time scale $1/\Omega_{\mathrm{K}}$. 
For simplicity, we assume instant cooling and heating, so we take the sound speed to be a step function, and choose $c_\mathrm{{s,min}}/c_\mathrm{{s,max}}$=0.6, as computed from the temperature in the shadows in our best radiative transfer model. 

\begin{figure}
	\centering
		\includegraphics[width=0.5\textwidth]{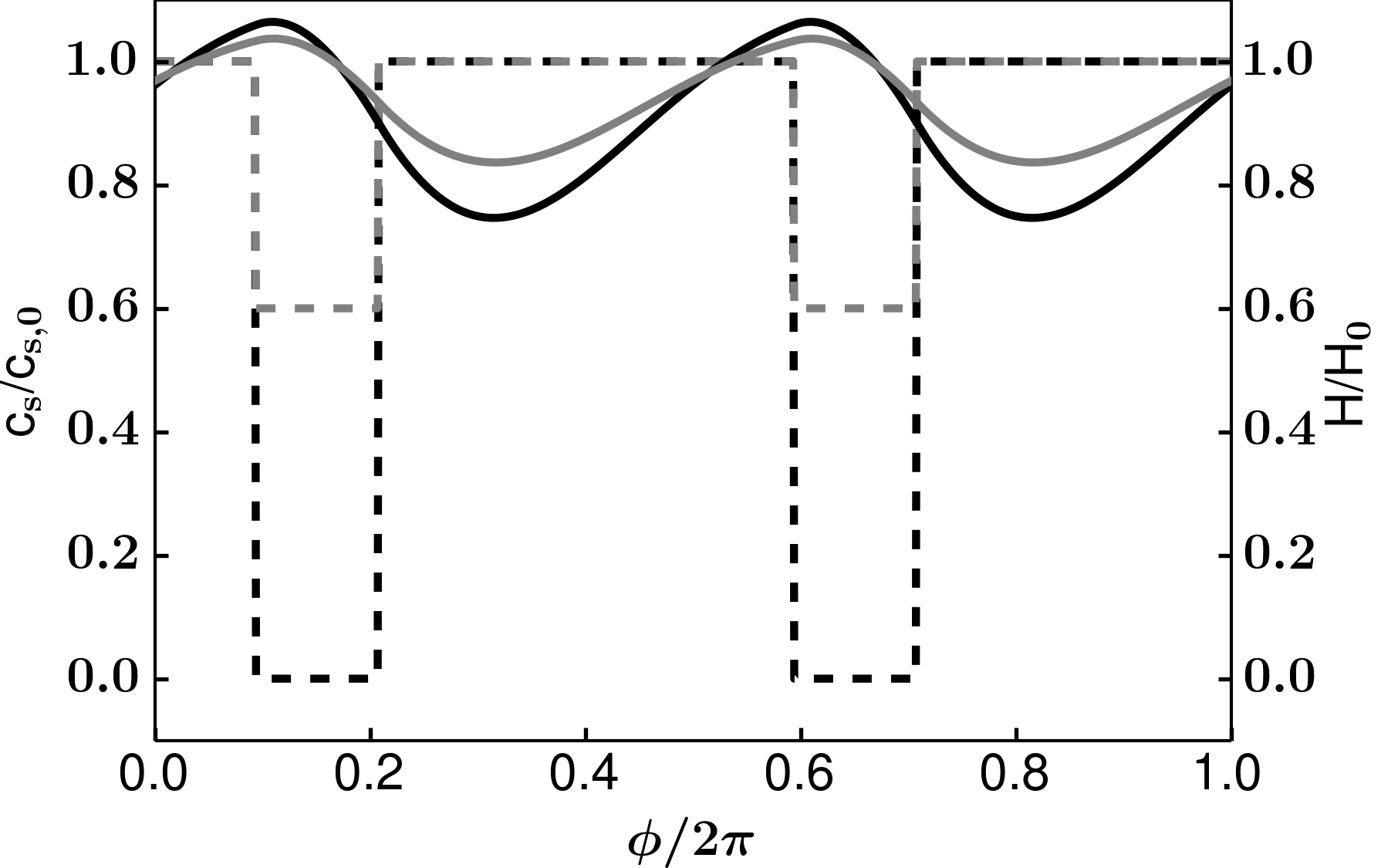}
	\caption{Isothermal sound speed profile (dashed) and scale height of the disk as a function of azimuthal angle along the ring (solid)  in two cases where $c_\mathrm{{s,min}}/c_\mathrm{{s,max}}$= 0.6 (gray) and $c_\mathrm{{s,min}}/c_\mathrm{{s,max}}$= 0 (black). All quantities are normalized. Note that the co-moving time increases towards the right. We set $H(t=0)=H_0=1.0$. }
	\label{fig:1d_results}
\end{figure}

Figure~\ref{fig:1d_results} shows the assumed sound speed profile and the modeled disk scale height for a single orbital period  (i.e., two periods in the oscillation because of the two shadows). Just before entering the shadow, the disk scale height reaches a peak height and increases above the initial value, due to the  inertia of the material.  A variation in scale height  changes the amount of stellar radiation intercepted by the disk and, at these locations, the rim  scatters more stellar light and appears brighter.  The width of this brightened region is related to the sound speed variation inside and outside of the shadows,  and to the damping parameter. This  leads to an asymmetric brightness distribution along the rim, the amplitude of which  is determined by the pressure difference between shadowed and illuminated regions.  Note that in Fig.~\ref{fig:1d_results}, the disk scale height is plotted against the azimuthal angle $\phi = \Omega_{\mathrm{K}} t$, which increases in the \textit{clockwise} direction to match the observed locations of the bright regions along the rim.  To approximately estimate the effect on the scattered light brightness, we assume that the brightness varies proportionally to the scale height, and multiply the scale height by the incoming radiation of the star, neglecting the effects of inclination and scattering angle. We find a maximum amplitude of 20\% brightness variation along the rim. In the extreme case of $c_\mathrm{{s,min}}/c_\mathrm{{s,max}}$=0, the maximum amplitude reaches 40\%, still significantly  less than the factor 2 observed (see Fig.~\ref{fig:cut}; between PAs of 125\dg\,and 270\dg, and PAs of 320\dg\,and 60\dg). 

In contrast, as shown in Sect.~\ref{sec:RT}, our radiative transfer model produces an azimuthally asymmetric brightness distribution that matches the observations well. This is due to the polarization efficiency being maximal along the semi-major axis.  This effect likely dominates, and can be amplified by the scale height variations along the rim, in particular on the far side of the (inclined) disk, for which we directly see the rim front. However, these scenarios cannot be disentangled because, by chance, the shadows are located close to the major axis. 


%
%
%
%

%

\subsection{Origin of a misaligned inner disk}
Shadows have now been detected in a handful of disks \citep[]{stolker2016, pinilla2015, canovas2016, avenhaus2014}.  A strongly misaligned inner disk is assumed to explain the presence of two shadows \citep{marino2015}, but the origin of such a misalignment is an open question. 

A massive planetary- or low stellar-mass companion that would carve a dust cavity inside 20\,au, and on an inclined orbit with respect to the outer disk, could possibly lead to a misaligned inner disk. Such a companion was detected in the cavity of the disk HD\,142527  \citep{biller2012, close2014} and found to be on an eccentric orbit \citep{lacour2016}.   If the outer disk holds a significant amount of gas, it is not clear how long such a misalignment can be sustained. Depending on the location and mass of the companion, the linear theory predicts that it can last \s 1 Myr at most \citep{foucart2013}. However, if the inner disk is highly misaligned, the timescale can be much longer due to the Kozai mechanism, an inclination/eccentricity pumping effect. If it is also on an inclined orbit, the M-dwarf companion could, in turn, influence the inner companion's orbit  \citep{lubow2016, martin2016}. 

 
 A massive companion inside the cavity could also explain the low gas-to-dust ratio and the very low mass accretion rate, estimated for this object \citep{collins2009}. The inner companion would halt material from flowing closer in towards the star, which would lead to an inner disk resembling a debris disk belt inside 1\,au. This inner belt should still be radially optically thick enough to cast two shadows on the rim, whilst having a scale height substantial enough to strongly reprocess light in the NIR regime. Dust at large scale height could be due to dynamical scattering of dust grains by the inner companion \citep{krijt2011}. 
    


\section{Conclusions}
In this paper, we present polarized scattered light optical and NIR images of the 10\,Myr  protoplanetary disk around the Herbig Ae star HD\,100453, obtained with SPHERE/VLT. We report on the detection of a ring like feature, two spiral arms, and two shadows located very close to the spirals. We also detect a faint spiral like feature in the SW.  

We present a radiative transfer model that efficiently accounts  for the main characteristics of these features, and discuss the hydrodynamical consequences of the change in stellar irradiation at the shadows' locations. We find that:
\begin{enumerate}
\item the properties of the shadows (location, width, contrast) are well reproduced using an inner and an outer disk misaligned by 72\dg. Their morphology depends on the inner disk aspect ratio, and on the width and shape of the outer disk rim;
\item the faint spiral-like feature detected in the SW could trace the scattering surface of the bottom side of the disk, if the disk is tidally truncated by the M-dwarf companion currently seen at a projected distance of 119\,au;
\item the strong azimuthal brightness variations observed along the rim can be well reproduced by the scattering phase function using small dust grains up to 1\,$\mu$m in size;
\item the local changes in stellar irradiation induces a modulation in the disk scale height that may amplify this effect. 
\end{enumerate}

The origin of the spirals, however, remains unclear. While the M-dwarf companion can produce the observed m=2 mode \citep{dong2016}, the clear connection of the spirals with the shadows is puzzling, and if not coincidental, means that the shadows may also play a role in triggering the spirals \citep{montesinos2016}. 
Another open question is how a 72\dg\,misalignment between the inner and outer disk can be generated, and whether this points towards the presence of an additional, yet undetected, massive companion inside the cavity. 

ALMA observations of this disk will undoubtedly shed light on many of these questions. It will not only be possible to estimate the gas and dust mass in the cavity and outer disk with more sensitive observations than the ones available today, but also to measure the kinematics of the gas. This may constrain the presence of a massive companion therein \citep{perezS2015}, and will indicate whether the spirals are leading or trailing, possibly constraining their formation mechanism. These observations will also accurately constrain the outer edge of the disk, which will then show whether the faint feature located in the SW is indeed the bottom side of a truncated disk, or is, in fact, another spiral arm. 

\begin{acknowledgements}
We acknowledge the team at Paranal for their help during the observations and the referee for his/her suggestions. M.B. wishes to thank J.-P.~Berger, M.~Flock, L.~Klarmann, and B.~Lazareff for fruitful discussions. 
SPHERE is an instrument designed and built by a consortium consisting of IPAG (Grenoble, France), MPIA (Heidelberg, Germany), LAM (Marseille, France), LESIA (Paris, France), Laboratoire Lagrange (Nice, France), INAF - Osservatorio di Padova (Italy), Observatoire de Gen\`eve (Switzerland), ETH Zurich (Switzerland), NOVA (Netherlands), ONERA (France) and ASTRON (Netherlands) in collaboration with ESO. SPHERE was funded by ESO, with additional contributions from CNRS (France), MPIA (Germany), INAF (Italy), FINES (Switzerland) and NOVA (Netherlands). SPHERE also received funding from the European Commission Sixth and Seventh Framework Programmes as part of the Optical Infrared Coordination Network for Astronomy (OPTICON) under grant number RII3-Ct-2004-001566 for FP6 (2004-2008), grant number 226604 for FP7 (2009-2012) and grant number 312430 for FP7 (2013-2016).  We acknowledge financial support from the Programme National de Plan\'etologie (PNP) and the Programme National de Physique Stellaire (PNPS) of CNRS-INSU. This work has also been supported by a grant from the French Labex OSUG@2020 (Investissements d'avenir, ANR10 LABX56). This work has made use of the SPHERE Data Centre, jointly operated by OSUG/IPAG (Grenoble), PYTHEAS/LAM (Marseille), OCA/Lagrange (Nice) and Observatoire de Paris/LESIA (Paris). The results reported herein benefitted from collaborations and/or information exchange within NASA's Nexus for Exoplanet System Science (NExSS) research coordination network sponsored by NASA's Science Mission Directorate.
AJ acknowledges the support by the DISCSIM project, grant agreement 341137 funded by the European Research Council under ERC-2013-ADG. A.Z., S.D, R.G. D.M., E.S.. acknowledge support from the ?Progetti Premiali? funding scheme of the Italian Ministry of Education, University, and Research.
\end{acknowledgements}

\bibliographystyle{aa}
\bibliography{hd100} 

\newpage
\appendix


\section{Angular differential imaging}
In this section we present angular differential imaging (ADI) images obtained with SPHERE in 2015 and in 2016. 

We reprocessed the 2015 data in the ESO archive that were published in \citet{wagner2015}. In the aforementioned discovery paper, reference differential imaging was used to investigate the inner structures of the disk (0.15-0.4\arcsec{}). This method outperforms ADI at the innermost radii (where large field rotation is required for efficient ADI), but changing conditions throughout the observations led to differences in the PSF of the reference star and science target and thus shallower than needed contrast to detect the fainter outer disk features. To recover these features in the 2015 data, we performed a second independent angular differential imaging reduction of these data, in which the intrinsic field rotation of the Alt-Az telescope is utilized to model the stellar PSF separately from the other astrophysical sources in the image. We post-processed the SPHERE-IFS data through analysis and subtraction of the principal components of the PSF via the KLIP method \citep{soummer2012} using self-developed IDL routines \citep{hanson2015,apai2016,wagner2016}. In modeling and subtracting the PSF from each science frame we rejected frames in which the field had rotated by less than 1.5$\times$FWHM pixel separation to avoid self-subtraction of the disk structures. Over the course of the observations the field rotated by 12.5\dg, allowing us to investigate the regions beyond 0.4\arcsec{} in high-contrast. The result is the detection at Y, J, and H-bands of the same faint third arm-like feature identified in the polarized intensity images, yielding confidence in its astrophysical nature. \\ 

In addition, \hd was observed on January 20th, 2016, as part of the SHINE survey for Guaranteed Time Observation (GTO), using the Dual Band Imaging mode \citep[DBI;][]{vigan2010dbi} of the IRDIS instrument, with dual band filters H2 and H3 simultaneously.  In parallel, a data cube was obtained with the near-IR Integral Field Spectrograph  \citep[IFS;][]{claudi2008} in YJ mode. These observations were obtained with the Apodized Lyot Coronagraph \citep[mask diameter: 185\,mas,][]{boccaletti2008}. We obtained a sequence of 4000 s in total on both instruments with a field rotation of 30 deg. Non-coronagraphic frames were obtained before and after the coronagraphic sequence for photometric calibration. Conditions were rather medium (seeing\s1.1\arcsec{}). The field orientation of IRDIS and IFS are derived from astrometric calibrations as described in \citet{maire2016}.  All the data were reduced with the SPHERE pipeline \citep{pavlov2008} implemented at the SPHERE Data Center together with additional tools developed for the handling GTO data reduction. This includes dark and sky subtraction, bad-pixels removal, flat-field correction, anamorphism correction \citep{maire2016}, and wavelength calibration for IFS. The location of the star is identified using the four symmetrical satellite spots generated by diffraction from a periodic waffle pattern introduced by  an appropriate modification of the adaptive optics reference slopes sent by the deformable mirror \citep{langlois2013}. Then, to remove the stellar halo and to achieve high contrast, the data were processed with the GTO high-level processing pipeline : SpeCal, which was developed for the SPHERE survey (R. Galicher, private communication).  

%


\begin{figure}[!h]
	\centering
		\includegraphics[width=0.45\textwidth]{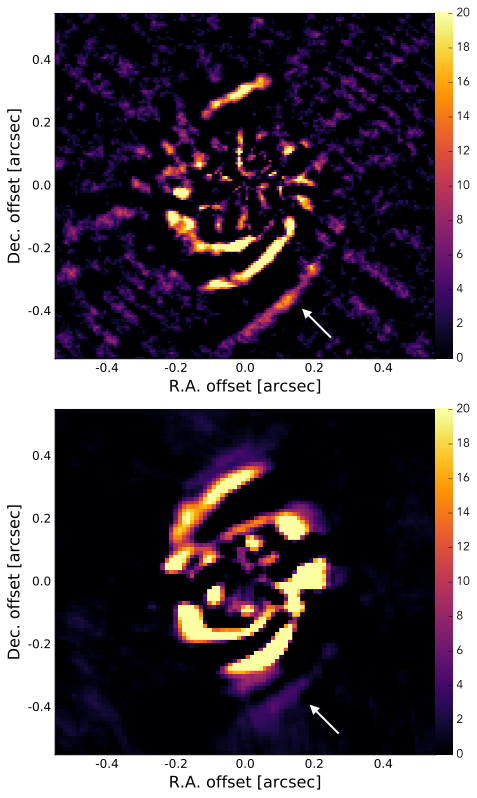} 
	\caption{ADI images from 2015 \citep[top, ][]{wagner2015} and from 2016 (bottom). The right arrows indicate the location of the faint third spiral-like feature that we interpret  as the outer edge of the scattering surface on the bottom side of the disk.}
	\label{fig:wagner}
\end{figure}

\section{RT modeling}
\begin{figure*}[!h]
	\centering
	\begin{tabular}{cc}
		\includegraphics[width=0.40\textwidth]{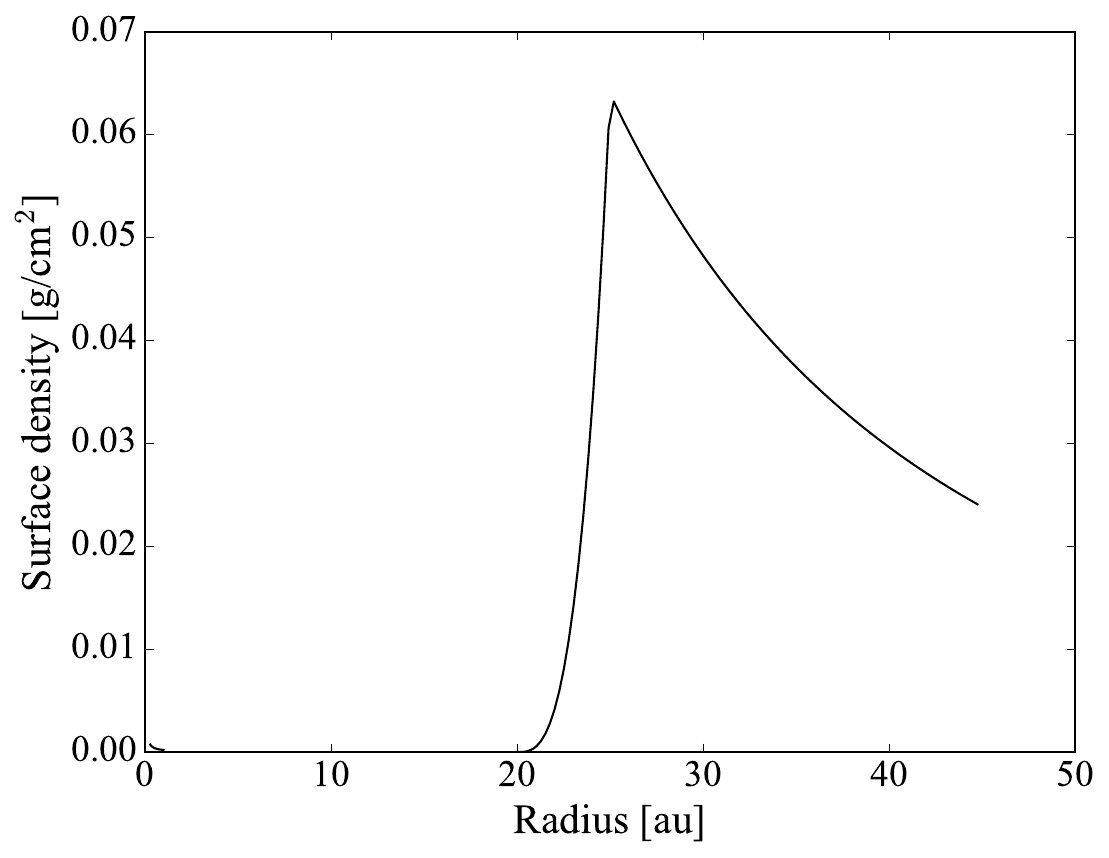} &	
		\includegraphics[width=0.40\textwidth]{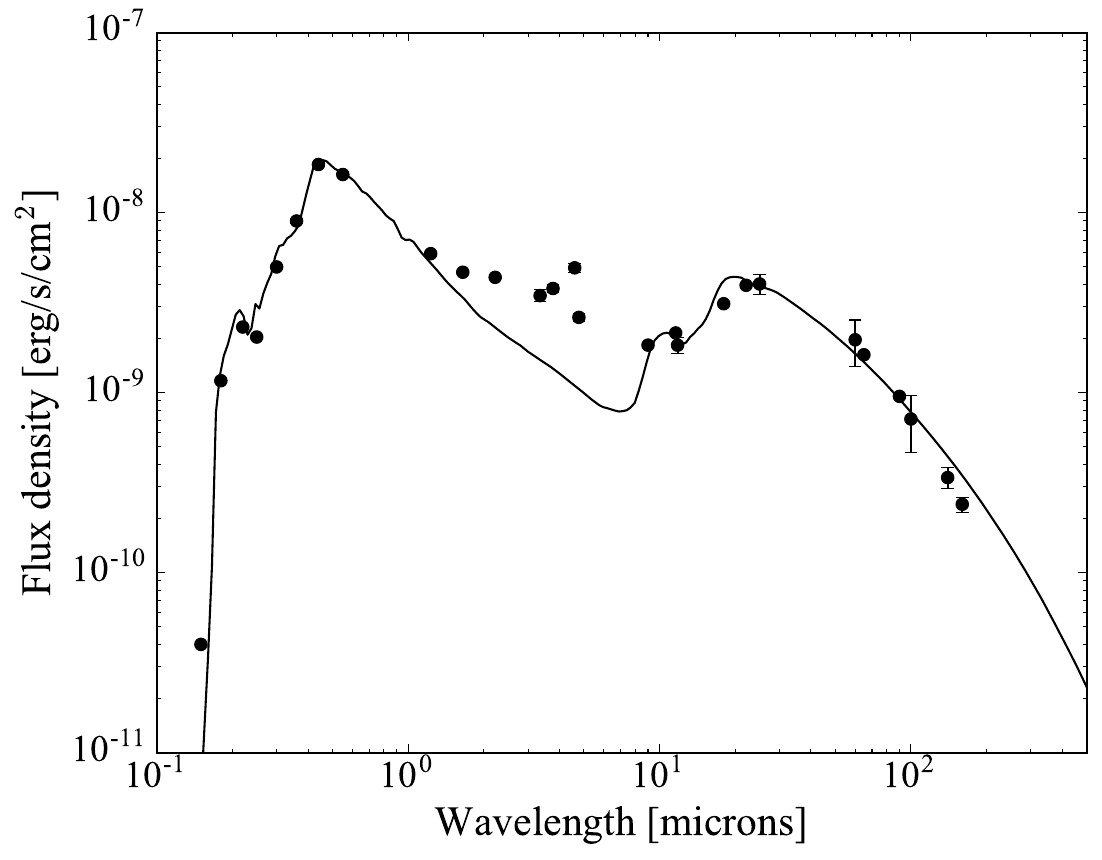}
	\end{tabular}
	\caption{Left: Surface density used in our radiative transfer model. Right: Modeled SED compared to the observed photometry (from \citet{khalafinejad2016}). }
	\label{fig:sigmased}
\end{figure*}
%

\end{document}